\documentclass[11pt,a4paper,dvipsnames]{amsart}
\usepackage{amsthm}
\usepackage{graphicx}
\usepackage{caption}
\usepackage{natbib}
\usepackage{subcaption}
\usepackage{amsfonts}
\usepackage{graphicx}
\usepackage{framed}
\usepackage{amssymb}
\usepackage{mathrsfs} 
\usepackage{tikz}
\usepackage{bm}

\usepackage{xcolor}
\definecolor{cornellred}{rgb}{0.7, 0.11, 0.11}

\usepackage[colorlinks=true,urlcolor=RoyalBlue,citecolor=RoyalBlue,linkcolor=RoyalBlue,linktocpage,pdfpagelabels,
bookmarksnumbered,bookmarksopen]{hyperref}

\usepackage{amsthm}
\usepackage{graphicx}
\usepackage{caption}
\usepackage{comment}
\usepackage{caption}

\usepackage{geometry}
\makeatletter
\@namedef{subjclassname@2020}{\textup{2020} Mathematics Subject Classification}
\makeatother

\numberwithin{equation}{section}
\title[Evidence Without Injustice]{Evidence Without Injustice: \\A New Counterfactual Test for Fair Algorithms}

\date{\today}

\begin{document}
\author[M.\,Loi]{Michele Loi}
\author[M.\,Di Bello]{Marcello Di Bello}
\author[N.\,Cangiotti]{Nicol\`o Cangiotti}

\address[M.\,Loi]{Algorithmwatch\newline\indent Linienstraße 13, 10178, Berlin, Germany}\email{michele.loi@icloud.com}

\address[M.\,Di Bello]{Arizona State University \newline\indent975 S. Myrtle Ave, Tempe, AZ 85287, United States} \email{mdibello@asu.edu}

\address[N.\,Cangiotti]{ Politecnico di Milano \newline\indent via Bonardi 9, Campus Leonardo, 20133, Milan, Italy}\email{nicolo.cangiotti@polimi.it}


\begin{abstract}
The growing philosophical literature on algorithmic fairness has examined statistical criteria such as equalized odds and calibration, causal and counterfactual approaches, and the role of structural and compounding injustices. Yet an important dimension has been overlooked: whether the evidential value of an algorithmic output itself depends on structural injustice. We contrast a predictive policing algorithm, which relies on historical crime data, with a camera-based system that records ongoing offenses, where both are designed to guide police deployment. In evaluating the moral acceptability of acting on a piece of evidence, we must ask not only whether the evidence is probative in the actual world, but also whether it would remain probative in nearby worlds without the relevant injustices. The predictive policing algorithm fails this test, but the camera-based system passes it. When evidence fails the test, it is  morally problematic to use it punitively, more so than evidence that passes the test. 
\end{abstract}

\maketitle

%% AI NOTE
\begingroup
\renewcommand\thefootnote{}\footnote{\textit{Disclaimer on AI use:} The authors used Claude 3.7 in the early stages of this project to help in the formulation of the Counterfactual Independence Principle presented in this paper. The model provided preliminary suggestions pertaining to the principle’s articulation, case studies, and the  structure of the argument. Screenshots of the initial output are on file with the authors. The AI-generated material was then analyzed, critiqued, and extensively revised by the authors throughout the writing process.}%
\addtocounter{footnote}{-1}%
\endgroup

\section{Introduction}
\label{Sec1}

Consider two policing algorithms and ask whether they are different from the point of view of fairness. The first analyzes historical crime data to predict where crime will occur and recommends allocating police patrols accordingly. Since historical data show that certain areas have higher rates of crime, the algorithm proactively assigns extra patrols to those areas. The second algorithm also allocates police patrols, but relies on cameras that are evenly distributed across the city. When a violent crime is reported, the algorithm scans nearby footage to identify suspects and directs officers to the scene or adjacent locations. 

Even though they work differently---one is predictive, the other diagnostic---both algorithms make assessments about crime locations and concentrate police in the same neighborhoods. Insofar as crime correlates with a neighborhood’s demographic composition (more on this shortly), the two algorithms will recommend dispatching police patrols more often to minority communities. As a result of the increased police presence, residents of minority neighborhoods will face a greater risk of being stopped, searched, arrested or killed by police, even when they did not partake in the crimes perpetrated in their neighborhood. The two algorithms, then, seem on a par from the point of view of (un)fairness. They end up allocating the harms of algorithmic error---false identifications, unnecessary stops, excessive force, fatal shootings---disproportionately on minorities. %But are they really on a par? 

%Is there any moral difference from the point of view of fairness between these two algorithms?  At first, it would seem there isn’t any. 

The realization that algorithmic error will fall disproportionately on minorities is likely to elicit different responses. Conservatives will point out that both algorithms track where crime is happening. They direct police to minority neighborhoods because that is where violent crime is concentrated, whether measured through historical crime data or camera recordings. The disparities in error rates across neighborhoods with different racial compositions are an inevitable, albeit unfortunate, byproduct of  crime patterns. Progressives will counter that this explanation does not say why crime clusters in this way. True, areas with larger minority populations have higher rates of violent crime.\footnote{By some estimates, the rate of violent crime in Black neighborhoods is, on average, five times higher than in White neighborhoods 
\citep{PetersonKrivo2010}.} 
But this clustering of crime is caused by historical processes that are morally suspect, such as segregation, redlining, and disinvestment, along with their negative effects on people’s well-being today, such as poverty, poor health and crime victimization.\footnote{For a causal analysis of the link between redlining and lower home ownership rates and property values, see \citep{Aaronson2021}. Other studies show a robust correlation between redlined census tracts and worse health outcomes \citep{Noelke2022etAl}. Longitudinal studies find that residents who grew up in redlined areas are at greater risk of violent victimization in adulthood, even after accounting for individual characteristics \citep{Poulson2021etAl,testa2024}. They are also at higher risk of being killed by police \citep{Mitchell2022}.}  Call these historical processes and their negative effects today \textit{structural injustices}.\footnote{According to a classic definition, structural injustice exists when `social processes put large categories of persons under systematic threat of domination or deprivation of the means to develop and exercise their capacities' \cite[p.~52]{young2011responsibility}.} 

%Once we acknowledge that the two algorithms operate against a backdrop of structural injustice---or perhaps despite that---our moral intuitions about their fairness start to diverge. For one thing, using camera footage to locate crime or decide where to dispatch officers strikes most of us as legitimate, even if the crime itself can be traced to structural injustice. In contrast, we find it troubling when police rely on historical statistical patterns, especially patterns that are due to unjust inequalities. The two algorithms---we will argue---are not on a par from the point of view fairness. A moral difference separates them, even though both track crime patterns; both distribute algorithmic harms unevenly across demographic groups; and both operate under conditions of structural injustice.\footnote{Some will also argue that both algorithms benefit minority residents the most, since increased police presence reduces crime rates and crime victimization \citep{RisseZeckhauser2004}.}

Once we acknowledge that the two algorithms operate against a backdrop of structural injustice---or perhaps despite that---our moral intuitions about their fairness start to diverge. Using camera footage to locate crime or decide where to dispatch officers does not seem unfair. Even if the crime itself can be traced to structural injustice, the camera merely reports an ongoing or very recent event, and the system recommends responding to it by dispatching police patrols to the scene.
To be sure, uneven placement of cameras across a city would allocate the burdens of surveillance unevenly and thereby raise fairness concerns. But the hypothetical we started out with assumed that cameras were evenly distributed. The predictive algorithm, by contrast, raises a more immediate fairness worry. It proactively sends police to areas where crime is expected to occur, and minority residents are more  likely to bear the costs of unnecessary stops, heightened scrutiny, or excessive force. This disparity is particularly troubling when the algorithmic forecasts rely on data about historical crime patterns that have been shaped by structural injustice. 

Admittedly, it is difficult to say exactly how, or even whether, the two algorithms differ from the standpoint of fairness. As seen earlier, both track crime patterns; both distribute algorithmic harms unevenly across demographic groups; and both operate under conditions of structural injustice.\footnote{Some will argue that both algorithms ultimately benefit minority residents, since increased police presence can reduce crime rates and victimization \citep{RisseZeckhauser2004}.}
In fact, the comparison between the two algorithms exposes a gap in current theories of algorithmic fairness. % We will examine more traditional statistical, counterfactual or causal criteria of algorithmic fairness; accounts based on compounding injustice and predictive justice; and finally accounts based on respects for individual agency and sensitivity to truth. 
We will argue that prevailing theories---whether statistical, counterfactual, causal, or focused on compounding injustice and predictive justice---cannot fully explain, or explain away, why the two algorithms seem morally different (Section \ref{Sec2}). We think that the difference lies in how structural injustice affects the evidential value of the data on which each algorithm relies. %Our proposed criterion, the Counterfactual Independence Principle (CIP), asks whether the evidential basis of an algorithmic output would retain its probative value in a world without structural injustice. 
For the predictive algorithm, the data have evidential value only in a world shaped by structural injustice. The same does not hold for the camera-based algorithm. That explains why the latter algorithm feels less unfair than the former (Section \ref{Sec3}).  We then extend this analysis beyond criminal justice to the health care domain (Section \ref{Sec4}).

A couple of observations before we begin. 
First, we do not claim that the camera-based algorithm is morally preferable in all respects. We only claim that, from the standpoint of fairness, it is less problematic than the predictive one. Our argument does not address separate questions about privacy, legitimacy, or community consent. Camera systems may well be more troubling on those fronts.  Second, notice that both algorithms  first identify crime locations, and as a second step, recommend punitive responses such as increasing police presence. But an algorithmic output  about a crime location could also be used to guide supportive responses, such as modifying the built environment to block criminogenic conditions. For clarity of analysis, however, we focus first on punitive responses and only consider supportive responses at the end of the paper (Section \ref{subsec:supportive}).

%To be clear, the difference in fairness between the two algorithms cannot lie in whether or not they use a race-based characteristic or a suspect's race to make assessments about the locations of crimes. It is true that a predictive policing tool that explicitly directed patrols to predominantly Black neighborhoods would be unacceptable. But the predictive policing algorithm under consideration relies on race   indirectly. It relies on crime data that is correlated with race. In contrast, the camera algorithm does use race explicitly. If footage captured “a Black man fleeing the scene,” the algorithm would incorporate that detail, along with other identifying features, to locate possible suspects. And yet this explicit use of race does not seem to raise a fairness worry.\footnote{Similarly, when a witness reports they saw a Black man near the crime scene, the identifying description `Black man' is considered useful information not to be ignored \citep{Banks2001Race-Bases-Susp}.} Paradoxically, the algorithm that explicitly uses racial descriptors feels less unfair.

\section{The Inadequacy of Existing Accounts}
\label{Sec2}

\subsection{Statistical and Causal Fairness}

Standard accounts of algorithmic fairness rely on statistical criteria, such as equalized odds or calibration across protected groups. An algorithm is fair by these metrics if its error rates, however defined, are equal across groups. Judged by these criteria, the algorithms in our two examples perform equally poorly: both produce different error rates across different racial groups. Insofar as police are disproportionately deployed to minority neighborhoods, whether because of historical crime data or video footage, these neighborhoods will experience higher false positive rates: more false identifications, unwarranted stop, arrests, and police shootings.  Statistical criteria, then, do not capture our different moral intuitions about the two algorithms.

But perhaps we should not dismiss statistical criteria so quickly. Insofar as predictive systems rely on crime data shaped by patterns of over-policing in minority neighborhoods, they will direct more patrols there and produce higher rates of false positives than camera systems, which track ongoing crimes more directly. In this sense, the two algorithms differ in their error rates. Suppose, however, that we adjusted the requirements for deploying patrols and made it more stringent whenever the evidence consisted of historical crime data. In this way, the false positive rates of predictive and camera-based systems would be equalized. Even under these conditions, the intuition of a moral difference between the two systems would not disappear. The persistence of this intuition is further reason to think that purely statistical criteria of fairness are inadequate.

In place of statistical criteria, some have proposed causal or counterfactual criteria. On the causal account, an algorithmic output is morally suspect if it is causally influenced by a protected attribute such as race \citep{chiappa_path-specific_2018}. On the counterfactual account, an algorithmic output is morally suspect if it would have been different had the individual’s race been different \citep{kusner_counterfactual_2017}. In both examples, however, race influences---or is counterfactually linked to---the algorithmic output. Either algorithm directs police to minority neighborhoods because those areas have higher concentrations of crime, and these concentrations are shaped by structural injustice, which is inseparable from race. In counterfactual terms, if the neighborhood’s racial makeup had been different, the pattern of disinvestment and crime rates would likely have been different too. Thus, either algorithm would be less likely to allocate extra patrols there. 

Causal accounts face another limitation: they treat race as a discrete, manipulable variable that can be turned “on or off”. The causal framing ignores how race is embedded in complex, structural phenomena \citep{hu_whats_2020}. If this is right, race cannot be reduced to a single variable, but is constitutive of the social conditions the algorithm responds to. We will discuss this objection in greater detail in the next section when we present our proposal. %, since we also deploy a causal and counterfactual analysis.

\subsection{Compounding Injustice}
\label{Sec22}
 Deborah Hellman's work on compounding injustice offers another lens through which to assess the fairness of algorithmic outputs \citep{Hellman23}. On her account, relying on evidence that reflects prior injustice is morally troubling because doing so will compound the original wrong. The predictive policing algorithm can be criticized for compounding injustice in Hellman’s sense. The first injustice lies in the historical conditions that have produced racially unequal patterns in crime data, such as housing policies and economic exclusion to the detriment of minority neighborhoods. The first injustice is compounded with another when the algorithm uses the tainted historical crime data to guide future police deployment. Communities already plagued by economic disadvantage, poorer health and higher crime victimization are then confronted with the harms of increased police presence. 

But---and this is the problem with Hellman’s account---the same argument applies to the video-processing algorithm. As before, the first injustice lies in the social conditions that give rise to racially disparate rates of crime. Segregation, underfunded schools, etc. have created environments where crime is more prevalent in minority communities. This injustice is  compounded with another when the algorithm, responding to specific crime incidents, disproportionately exposes innocent individuals from minority communities to unnecessary stops, searches and fatal police encounters. The harms of policing, even when based on incident-specific evidence, fall unevenly on already disadvantaged groups. 

\subsection{Predictive Justice}

We now turn to a third attempt. \cite{LazarStone} have proposed this principle for evaluating predictive algorithms: 
\begin{quote}
\textit{Predictive Parity.} An algorithm is predictively just only if its performance for disadvantaged groups cannot be improved without a disproportionate decline in performance for advantaged groups. 
\end{quote}
%In motivating this principle, 
Lazar and Stone emphasize that the moral significance of a performance disparity---say a difference in the rates of false positives for two groups---depends on the degree of background injustice affecting the disadvantaged group. For suppose the same predictive model is deployed in two  societies, and in both the model makes fewer errors for the advantaged than the disadvantaged group. If the disadvantaged group in the second society suffers from greater structural injustice than the one in the first, the same performance disparity should be more morally troubling in the second society. The deeper the background injustice, the more urgent the moral imperative to redress the disparity. 

A related point concerns the cause of the group disparity in algorithmic performance. If a performance gap stems from structural injustice---for example, from disinvestment in certain neighborhoods that leads to higher rates of poverty and crime, which in turn leads algorithms to predict more crime will occur in those areas---the disparity carries greater moral weight. It generates a stronger obligation to mitigate the performance gap than a disparity resulting from morally neutral factors, such as differences in behavior attributable to lifestyle choices.

We agree that performance disparities often reflect structural injustice and that the social context in which a predictive algorithm is deployed matters for its moral evaluation. The Predictive Parity Principle, however, does not explain the moral difference between the two algorithms in our examples, since both are operating within the same structurally unjust society.  %and the performance disparity is caused by injustice. %Perhaps the principle could be interpreted to distinguish between them---for example, based on the nature of the harms involved---but this is far from clear without guidance on how to operationalize the principle in practice.

\subsection{Agency}

We conclude this section by discussing two other promising, though ultimately inadequate, accounts of the moral difference between the predictive policing algorithm and the video-processing algorithm. These accounts have not been explicitly discussed in the literature on algorithmic fairness. They are found in the literature on the uses of statistical evidence in legal proceedings. 

The first invokes the idea of agency \citep{Wasserman91, pundik2017}. One might argue that predicting policing fails to respect individual agency, whereas the video-processing algorithm that responds to specific crime incidents does not fall prey to the same problem. On this view, the moral difference lies in whether or not individuals are treated as autonomous agents endowed with deliberative capacities. The second algorithm appears unproblematic in this regard, as it responds to  evidence about crimes that have already occurred and does not make forward-looking predictions about people’s behavior. Instead,  predictive policing seems to conflict with respect for agency, since it relies on historical patterns to anticipate where future crimes will take place. 

But reliance on predictions about human behavior does not automatically preclude respect for human agency. Human behavior is predictable because it is influenced by environmental, psychological, or character-based factors. This influence does not eliminate agency unless one subscribes to a libertarian view of free will. In addition, the predictive policing algorithm makes neighborhood-level predictions. It does not assess individual culpability or predict individual actions, but instead recommends where to deploy police resources based on historical crime data. The unit of prediction is the neighborhood, not the person. As such, any tension with individual agency is indirect and limited. 

Another difficulty here is that the agency account fails to explain our different moral intuitions in cases in which individual behavior, not just group-level patterns, is treated as predictable. Consider two tools used to assess recidivism risk. The first relies on demographic and neighborhood characteristics, such income and ZIP code. The second draws on physiological stress markers such as cortisol levels and heart rate variability, which are, let us suppose, reliable indicators of a more impulsive temperament. They are biological indicators beyond conscious control. Both tools treat individual behavior as the output of measurable predictors, and both bypass people’s deliberative capacities. Yet many will find the physiological approach less morally troubling than demographic profiling. %This intuition suggests that what matters is not only whether a predictive tool disregards agency, but also which predictors it relies on and how they are socially understood. Demographic variables like race and neighborhood are entangled with structural injustice, while physiological indicators are less socially loaded. The agency account fails to accommodate this nuance. 

\subsection{Tracking truth}
\label{Sec25}
The final attempt we consider for distinguishing the two algorithms rests on the epistemological point that the camera algorithm is preferable because it better tracks the truth of ongoing crimes \citep{enoch_statistical_2012}. The idea of tracking the truth is often formulated counterfactually.  If no crime had occurred in a given area, the camera would detect nothing and no officers would be dispatched. In contrast, the predictive policing algorithm would still predict crime in the area no matter what, since it relies on historical crime data. 

%There is some force to this point. 
This argument draws attention to the fact that one system is predictive and the other diagnostic. Predictive algorithms forecast where future crimes will occur; diagnostic ones identify crimes that already happened or that are ongoing. Predictive policing is often viewed as riskier because it can create feedback loops that are detached from actual trends in crime: more patrols lead to more recorded crime, which then justifies more patrols \cite[Chp.~5]{harcourt2007}.
%\footnote{ \red{SISTEMARE} The net effect is that minority neighbourhoods are disproportionately represented in police encounters compared to their actual crime rates.If A group offends at a 6\% rate while the other B group offends at 4\% rate (so 6:4 ratio), a proportional representation in police encounters would be 60\%/40\% for the two groups (again 6:4 ratio). Harcourt worked out this critique in his 2007 book Against Prediction: Profiling, Policing, and Punishing in an Actuarial Age. See, in particular, Chapter 5 on the Ratchet effect (\href{https://www.tau.ac.il/law/events/15-12-06/C-Chapter5.pdf}{https://www.tau.ac.il/law/events/15-12-06/C-Chapter5.pdf}) (“[W]hen the police profile higher-offending individuals, they are effectively sampling more from that higheroffending group. The resulting set of successful searches will contain a disproportionate number of those high-offending individuals—disproportionate as compared to their representation in the offending population. This imbalance will get incrementally worse each year if law enforcement departments rely on the evidence of last year’s correctional traces—arrest or conviction rates—in order to set next year’s profiling targets.”) cited by The Challenges of Prediction: Lessons from Criminal Justice p. 161.} 

But not all predictive policing is objectionable. It is reasonable to deploy officers at large events, busy transit hubs, or during protests. These are places where crime is more likely. Prediction is essential to policing, even though had crime not happened in those areas the police would have been dispatched there anyway. A blanket distinction between predictive and diagnostic tools is too crude to capture what is morally at stake.

\section{Counterfactual Independence}
\label{Sec3}

\subsection{Counterfactual Independence Principle (CIP)}

The shortcomings of existing accounts suggest the need for a novel principle. As a starting point, observe that the predictive policing algorithm and the video-processing algorithm rely on different forms of evidence to make assessments about locations of crime. One draws on historical crime data, often correlated with neighborhood demographics. The other analyzes the footage of specific incidents to identify criminal activity and potential suspects.  These forms of evidence---we will argue---differ in how their probative value depends on structural injustice. Even in a society free from structural injustice, cameras would still capture perpetrators at crime scenes. But predictive policing would lose its truth-tracking power. 

This observation points to the following principle: in assessing the moral acceptability of deploying an evidence-based algorithm, we must ask not only whether the evidence it relies on is probative of the outcome of interest, but also whether the evidence would remain probative absent structural injustice. %f the evidence derives its probative value from unjust social arrangements,  this fact itself is morally relevant in deciding how the evidence should be used downstream. 
%For example, when considering whether to dispatch police to neighborhoods flagged as high-crime, we should take into account whether the evidential link between neighborhood and crime would persist in a more just society, rather than simply acting on the evidence because it is probative. %This section is devoted to articulating this principle in more detail. 
%
A more formal statement of the principle can be broken down into two parts:
\begin{quote}
\textit{CIP Test}. For evidence $E$ about outcome $O$ (e.g., the location of crime), check whether $E$ would retain its probative value about $O$ in nearby possible worlds where the relevant structural injustices are absent. If yes, $E$ passes the test, or else it fails it. 
\end{quote}
\begin{quote}
\textit{Downstream Use}. Whether $E$ passes the CIP test is a morally relevant consideration for determining its acceptable downstream uses. For punitive uses of the evidence (e.g., dispatching police or restricting personal liberties), we should exercise greater caution when the evidence fails the CIP test than when it passes it. 

\noindent
(For non-punitive, supportive uses, see Section \ref{Sec4}.)
%Whether $E$ passes the CIP test is a morally relevant factor in deciding about acceptable downstream uses of the evidence. When the evidence is used punitively (e.g., to dispatch police and restrict personal liberties) we should exercise greater caution in case the evidence does not pass the CIP test compared to evidence that does. (For non-punitive uses, see Section \ref{Sec5}.)
\end{quote}
%
%Evidence $E$ about an outcome $O$---say, the location of crime---is morally preferable to evidence $E^*$ about $O$ if (i) $E$ would retain its evidentiary value about $O$ in nearby possible worlds without relevant structural injustices, whereas $E^*$ would lose most or all of its value in such worlds, and (ii) the evidence in question is intended to be used in a punitive way. 
%
%We will examine the second part of the principle in greater detail later in the paper. For now, it is important to note that evidence---and the algorithms that rely on it---can be deployed in both punitive and supportive ways. When the use is punitive, the principle directs us to exercise particular caution if the evidence fails the CIP test. In these cases, the fact that the evidence derives its probative value from structural injustice makes its use especially problematic, far more so than if the evidence had passed the test. We will consider supportive uses later. 

The principle helps to explain our different moral intuitions about the two algorithms. Take the video-processing system. Its recordings will contain identifying features such as clothing, gait, and other distinctive characteristics. Some of these---like facial structure, skin tone, or hair texture---will correlate with racial identity. But their evidentiary value derives from a presumed causal connection to the events of the crime, or to events that precede or follow it. That causal link would exist even in the absence of injustice. The value of video evidence, and of the identifying characteristics it reveals, would thus remain intact in a just society. It does not depend on structural injustice for its relevance or reliability. 

In the predictive policing example, the accuracy of the algorithm's forecast crucially depends on the systematic spatial concentration of crime, which itself is a legacy of structural injustice. Without segregated housing, economic exclusion, and targeted disinvestment, crime would likely be more evenly distributed. Without the structural conditions that make neighborhoods criminogenic, past crime locations would be poor indicators of future crime locations.  

There are, of course, exceptions to the claim that crime would not cluster geographically in the absence of structural injustice. One example is an entertainment district filled with bars, music venues, and late-night restaurants. People voluntarily gather there, often carrying cash, drinking, and staying out late. These conditions increase the likelihood of thefts, assaults, or disturbances in general. Or take a busy station. Large crowds and moments of distraction create favorable conditions for pickpocketing and other petty crimes. Crime clusters because of patterns of use and behavior, not because of structural injustice. 

The Counterfactual Independence Principle (CIP) does not deny that historical crime data, in some cases, could retain predictive value in a just society.
A predictive policing algorithm that relied on factors such as a busy train station or a entertainment area would continue to track crime accurately even absent structural injustice. By CIP, and in line with our moral intuitions, such an algorithm would be no more unfair than one that relied on video evidence. %And This verdict aligns with our moral intuitions. 

% These examples clarify the contours of the distinction we are trying to draw. When a predictive algorithm relies on evidence that remains valuable even without injustice, the moral distinction between it and video-processing weakens. If both retain their epistemic value in a just world, the moral preference for one over the other diminishes accordingly. . %It alerts us to whether the value of evidence is contingent on injustice. 

\subsection{The CIP test}

Since cases cannot be neatly categorized, careful attention to nuance is essential in applying the first part of the principle, the CIP test. In practice, this means examining the causal mechanism that make a piece of evidence probative and asking whether those mechanisms would still hold in the absence of structural injustice. Here is a step-by-step heuristics:
\begin{itemize}
    \item[I.] identify the mechanism that makes evidence $E$ relevant to an outcome $O$;
    \item[II.] determine whether the mechanism is part of structural injustice;
    \item[III.] if it is, evaluate a nearby counterfactual world where that unjust mechanism is removed, and ask whether the evidentiary link still holds; 
    \item[IV.] if it does, $E$ passes the CIP test, or else it fails it. 
\end{itemize}
Consider again our running example of predictive policing. Suppose there is a well-supported and stable causal explanation for why minority neighborhoods have higher rates of violent crime. The causal chain might be this: 
\begin{quote}
Minority neighborhood $\rightarrow$ Redlining $\rightarrow$ Segregation $\rightarrow$ Neighborhood Disinvestment $\rightarrow$ Underfunded Schools $\rightarrow$ Limited Employment $\rightarrow$ Concentrated Poverty $\rightarrow$ More Crime.
\end{quote}
There is no doubt, given this causal mechanism, that historical crime data is predictive of crime. But several links in the causal chain---for example, redlining and segregation---reflect policies and practices now recognized as unjust. In a just society where redlining and segregation had never occurred, the causal chain would be broken, and the correlation between location and crime would likely not exist. Since the predictive value of historical crime data depends---at least in part---on unjust policies, this evidence fails the CIP test. 

Now consider a different causal mechanism. Busy areas---like train stations or entertainment districts---attract crime because they draw crowds, distractions, and opportunities for theft or disorder. The causal chain might be: 
\begin{quote}
Busy location $\rightarrow$ High Density of People $\rightarrow$ Anonymity and Distraction $\rightarrow$ More Crime.
\end{quote}
The causal pathway in this case does not contain any policies or processes we regard as unjust. So the evidence here---historic crime data---passes the CIP tests since it would retain its probative value even when unjust social structures are removed. The same goes for video evidence, which rely on a direct link between the event and the captured image:
\begin{quote}
Crime in minority neighborhood $\rightarrow$ Camera Captures Perpetrator $\rightarrow$ Crime location information. 
\end{quote}
These forms of evidence are diagnostic and event-specific. Their evidential value does not depend on the broader social context in which they occur.

\subsection{Comparison with causal and counterfactual accounts}

Our approach relies on causal and counterfactual judgments about the probative value of the evidence, but differs from mainstream causal and counterfactual accounts of algorithmic fairness  (see Section \ref{Sec2}). The latter focus on whether an algorithm's prediction would change if an individual's race were changed, holding all other relevant variables fixed. Fairness is assessed by intervening on the variable `race' and tracing the effects of this intervention on the algorithm’s output. If race has a causal impact on the output, the algorithm is considered unfair.\footnote{Some causal accounts distinguish between morally permissible and morally impermissible causal pathways involving race \citep{chiappa_path-specific_2018}. Consider an algorithm that uses ZIP code as a predictor of crime. While race does not directly determine someone's ZIP code, there is often a causal connection between race and residence due to structural forces like redlining. In this sense, race causally influences someone’s ZIP code through the history of segregation. If the algorithm relies on ZIP code, its output is indirectly affected by race through a morally problematic pathway. By contrast, consider an algorithm that uses a medical biomarker that correlates with race due to genetic variation, and that biomarker is causally predictive of a health outcome. In such a case, race is still affecting the algorithmic assessment about someone’s health but it is not part of a morally impermissible causal pathway.} 

The problem with the causal or counterfactual approach, as already noted, is that they cannot explain the moral difference between the camera-based algorithm and the predictive policing algorithm. In both cases, race causally influences the algorithm's output and the causal pathways that contains race are shaped by structural injustice. Even in the case of video-based algorithms, which are diagnostic and specific, the camera captures crime where crime is concentrated. But the concentration of crime reflects a causal chain rooted in structural injustice: racial segregation, disinvestment, poor schooling, etc. The evidence---the location flagged---is influenced by these unjust background conditions.

%\subsection{Hu and Kohler-Hausmann's Critique}

%Another difference between our approach and causal and counterfactual approaches lies in their susceptibility to an influential critique. 

An additional problem is that causal and counterfactual approaches treat race as a variable that can be manipulated independently of everything else, but in a racially stratified society, this assumption of independence breaks down \citep{hu_whats_2020}. Race is entangled with everything else: where someone lives; the school they attend; the jobs they can access; etc. The question “Would the algorithm output have been different if this person were of a different race, holding everything else constant?” doesn’t make sense. Changing race while keeping everything else fixed cannot be done. 

The CIP test avoids this problem because it shifts the counterfactual question from ``Would the algorithmic output still be the same if we changed $R$?" to ``Would $E$ still be evidence for $O$ if we changed the unjust mechanisms that connected $E$ and $O$?" In asking whether the algorithm output would have been the same if the person’s race $R$ were different, race needs to be isolated from other variables. Hu and Kohler-Hausmann's critique then applies. By contrast, CIP asks about the epistemic relevance of a relation: whether a piece of information $E$ (such as a person’s neighborhood) would still be good evidence for some outcome $O$ (such as the location of crime) in a world without structural injustice. Here the counterfactual requires to isolate and manipulate the unjust social processes, not race.  The critique no longer applies.

\subsection{Isolating unjust social processes}

But how can unjust social processes be manipulated? %The standard rule in evaluating counterfactuals is to minimize the deviation from the actual world: we change what is necessary to make the antecedent true, but leave all else unchanged unless required by causal consistency.  In other words, we consider nearby possible worlds, that is, worlds as similar as possible to the actual one, except for the specific change we are interested in. 
We counterfactually hypothesize a world without certain structural injustices (e.g., without redlining), and then ask whether the evidentiary connection between $E$ and $O$ would still hold in that world. One might extend Hu and Kohler-Hausmann's critique to argue that any attempt to remove unjust social structures in a counterfactual world while holding other things constant is incoherent, because unjust structures affect everything. Just as we cannot ``turn race off" without collapsing the rest of the causal graph, we cannot imagine removing redlining or segregation without unraveling the entire social fabric. 

This objection goes too far. It amounts to claiming that no causal relation can be meaningfully evaluated apart from structural injustice. Many processes in the natural and social world operate independently of historical oppression or discriminatory policy. Mechanistic causal systems---such as the physics of camera operation, the biochemical process by which a drug affects the body, or the structural integrity of a bridge---are examples of systems whose causal functioning does not depend on structural injustice. A camera's ability to capture movement, light, and spatial location is governed by physical laws and engineering design. While the placement of the camera may be influenced by patterns of policing or surveillance shaped by injustice, the mechanism that links a person’s physical presence at the scene to the resulting image is not.\footnote{We can distinguish five types of mechanisms by which evidence can gain probative value: 1. physical mechanisms, such as camera optics; 2. biological mechanisms, for instance, how drugs are metabolized in the body; 3. behavioral patterns, such as crowd dynamics; 4. social mechanisms that are independent of justice, such as the operation of supply and demand in markets; and 5. social mechanisms that depend on injustice, such as racial residential segregation.}

%While social injustice may reach further than we think, it does not reach everywhere. It is reasonable to isolate and remove unjust causal pathways while leaving other parts of the causal model intact. The CIP framework invites us to test whether a particular evidentiary link---$E$ supports $O$---relies on mechanisms that we have a moral reason to be suspicious of.
Another way to press the objection is to claim that we cannot know what would happen in a world without structural injustice. How can we assess whether evidence E would still support outcome $O$ if we removed redlining, racial segregation, or concentrated poverty? Since these structural conditions have shaped the world so deeply, imagining their absence will be speculative. If so, the CIP test would rest on shaky epistemic ground. It would ask us to evaluate evidentiary relations in worlds we cannot empirically access with confidence. 

This epistemic limitation is real, but not fatal. We always face uncertainty whenever we evaluate counterfactuals, but we don't need infallible knowledge of the counterfactual world, only reasonable confidence based on the available evidence and applicable causal theories. Consider the question: Would eyewitnesses still observe and report suspects fleeing a crime scene in a just world? The answer is almost certainly yes. The causal connection between perception, memory, and testimony is not dependent on redlining or racial exclusion. Compare that to the question: Would a person’s ZIP code still predict their creditworthiness in a just world without residential segregation or racial stratification? Likely not. ZIP codes are predictive today because they encode patterns of racialized access to wealth. If those background injustices were removed, the evidentiary link between ZIP code and credit risk would collapse. 

\section{Applications of the Principle}
\label{Sec4}

\subsection{Criminal Justice}
The Counterfactual Independence Principle (CIP) provides a useful lens for distinguishing between evidence that is morally acceptable in criminal justice and evidence that is tainted by structural injustice. Traditional forms of evidence---such as DNA samples, fingerprints, and video footage---are generally permissible under the principle because their evidentiary value rests on causal connections to the crime that do not depend on unjust social conditions. 

What about character evidence and prior criminal history? Will they fail the CIP test? The answer ultimately depends on why prior offending is predictive of future criminal behavior. If people who have offended in the past continue to be exposed to social and environmental pressures that are criminogenic---pressures that result from structural injustice---then the evidentiary value of prior crimes would be objectionable under CIP. If this is right, absent structural injustice, prior criminal behavior would lose its evidentiary value. In addition, when we speak of “prior crimes,” we often mean prior arrests or convictions. These records are themselves shaped by patterns of discriminatory policing and prosecution. In a society without racial discrimination, the correlation between prior convictions and future crime could well disappear. %This would undermine the reliability of prior crime evidence.

The same logic extends to evidence that is used by algorithmic tools in criminal justice. Algorithmic tools that process permissible evidence---for instance, matching DNA samples or authenticating surveillance camera footage---pass the CIP test because they do not rely on unjust social structures. By contrast, predictive policing systems that rely on historical neighborhood crime patterns, such as PredPol or HunchLab, are problematic. As seen before, these tools draw their predictive power from long-standing spatial clusters of crime that exist because of structural injustices like segregation and economic disinvestment. Predictive policing need not always be problematic, however. A predictive tool that responds to immediate, short-term repeat victimization---such as a rash of break-ins on a single block over several nights---should be acceptable under CIP. Here, the predictive inference stems from a direct behavioral pattern that would still be relevant in absence of structural injustice. 

Risk assessment tools like COMPAS or the Public Safety Assessment (PSA) can be assessed in a similar manner. These tools predict a defendant’s risk of reoffending based on factors such as prior arrests, address of residence, or employment status---variables that directly reflect structural disadvantages. Their predictive accuracy depends on patterns that would likely dissolve in a structurally just society. But, as before, we can imagine a risk assessment tool that meets the CIP test. For example, an instrument that relies only on direct, case-specific indicators---such as the nature and severity of the current offense, credible threats made during the crime, or statements indicating an intention to flee---could retain its probative value regardless of social background conditions. %The key is that the inference should not be grounded in structurally unjust conditions. 

\subsection{Healthcare}

Healthcare offers other illustrative cases. %\footnote{For a general discussion algorithmic fairness in medicine and healthcare, mostly focused on performance fairness criteria, see Algorithm fairness in artificial intelligence for medicine and healthcare - PMC.} 
Consider algorithms that estimate the likelihood that a patient will benefit from a treatment or drug that is costly or scarce. It is instructive to distinguish three different scenarios. First, suppose a patient’s reduced ability to benefit from a drug is due to genetic factors, for instance, a genetic variant interferes with how the drug is metabolized and reduces its effectiveness. If an algorithm used this information to guide treatment allocation, it would satisfy the CIP test: the predictor---genetic variation---would still be relevant in a just world. Allocating scarce resources based on biological differences is generally considered appropriate.\footnote{The medical research community is not off the hook: there remains a duty to develop equally effective treatments for patients with less favorable genetic profiles.} 
%Without such efforts, what begins as a biologically justified distinction can lead to a sustained and unjust disparity in health outcomes, simply because one group happens to be excluded from the benefits of medical innovation.}

The second case is analogous to the camera evidence case. Suppose a comorbidity, say type 2 diabetes, reduces the effectiveness of a drug, but having it is itself partly due to structural injustice.  Certain comorbidities are more prevalent in marginalized communities due to long-standing disparities in access to nutritious food, preventive care, clean environments, etc. If the drug biologically works less well in patients with type 2 diabetes, this biological relationship would still hold in a just world, even if the comorbidity will be less prevalent among disadvantaged populations. Thus, an algorithm that relied on the comorbidity as predictor would pass the CIP test. %This case is analogous to the camera algorithm in policing. %
%But, unlike the purely biological case, a pressing moral concern here arises. If treatment is systematically limited for those less likely to benefit---when the reason they benefit less is that they were made sicker by injustice---the allocation compounds an existing disadvantage. Ideally, 
Policy interventions should address the root causes, but the algorithm allocating a scarce treatment cannot correct for that. Ignoring differences in the likelihood of treatment success would not help those most disadvantaged since they do have lower chances of therapeutic benefit. 

Here is a third scenario. To guide the allocation of treatment, a predictive algorithm relies on social variables such as lower income and housing instability. These factors are strongly correlated with poor treatment outcomes, not because of any biological limitation, but because of difficulty attending follow-up appointments or storing medications. Although lower income and housing instability are predictive, they fail the CIP test: in a fairer world with equal access, they would no longer predict reduced treatment effectiveness. Using such predictors to limit treatment would be less morally defensible than in the second scenario. Here, patients are denied care not because the drug is less effective on their bodies, but because unjust social conditions will interfere with its administration. %The CIP test captures this distinction: it fails here but succeeds in the second scenario. 

In the second and third scenario, injustice is compounded: denying already disadvantaged individuals access to medical treatment further entrenches their disadvantage. But there is a meaningful moral difference. In diabetes case, the reduced effectiveness is biologically real, even if socially caused; in the third case, the barrier is almost entirely logistical. The CIP test detects this difference.

Another moral---and political---difference is that a downstream correction will only work in the third scenario. Instead of restricting treatment for disadvantaged patients based on lower predicted success, a more inclusive allocation strategy would correct the social barriers that reduce effectiveness. Disadvantaged patients can be given additional support---such as transportation, housing assistance, or case management---to ensure that the treatment is just as effective for them as for others. 
In the second scenario, however, such downstream correction is not available. Patients with certain comorbidities are genuinely less likely to benefit from the drug, and no amount of logistical or social support can change that. The reduced efficacy is rooted in physiological factors, even if those factors have unjust origins. The upstream causes must be addressed, but no adjustment at the point of allocation can make the drug work better for these patients. 

\subsection{Not only punitive responses}
\label{subsec:supportive}

This discussion reveals an interesting reversal. When the response to an algorithmic prediction or assessment—say, of drug effectiveness—is punitive, failure to meet the CIP test counsels against that punitive response. But when the response is supportive, failure to meet the CIP test seems to provide an additional reason in favor of the supportive intervention. %Precisely because disadvantaged patients face barriers to adhering to treatment, supportive responses are especially warranted. 
%Although our formulation of CIP has been limited to algorithmic outputs that are used punitively---to burden, restrict, or punish individuals---we should now examine how CIP would handle cases in which the algorithm outputs are used supportively, for example, to allocate resources, deliver preventive interventions, or direct services. 
%
If patients are unable to adhere to treatment because of barriers such as lack of access, resources, or transportation, the appropriate response is not to block them from treatment on grounds of lower predicted effectiveness, but to design supportive interventions that remove those barriers. %In this way, lower treatment effectiveness becomes a reason in favor of providing additional support, not for restricting care. 

Similarly, consider an algorithm designed to identify patients most likely to benefit from intensive diabetes prevention programs. The algorithm relies on historical data that shows that individuals in certain minority communities face elevated risks of developing diabetes, in part because of historical and ongoing structural injustices. Such evidence  fails the CIP test, but it does not follow that the evidence should be set aside. On the contrary, precisely because the higher risk of diabetes arises from structural injustice, the case for using the evidence to direct preventive care becomes stronger. %Here failing the CIP test is not a red flag against using the data, but a caution against neglecting the data. 
Here, the fact that the evidence rests on structural injustice is not a reason to disregard it. On the contrary, it strengthens the case for using the evidence as a guide to supportive interventions—both immediate and long-term—that mitigate, rather than reproduce, structural harms.

Policing presents a similar case. If historical crime data show that certain neighborhoods have higher rates of crime, this information can be used punitively, for instance, by sending more patrols, expanding stop-and-frisk practices, or justifying surveillance. But the same evidence could instead guide supportive interventions. City officials might install better street lighting, provide neighborhood watch support, or fund community mediation programs to defuse local tensions. These measures reduce opportunities for crime and build trust without increasing punitive surveillance.\footnote{One reason for hesitation here is the belief that supportive social interventions are less effective at reducing crime than increasing police presence. This may be so. But police presence itself can be framed as either punitive or supportive. Officers who operate as fellow neighbors---they are around to help residents---can be seen as supportive. Instead, when police function as an external force of surveillance and control, their presence takes on a punitive character. If we conceive of policing as supportive in this way, we would also be less inclined to object to the use of crime data that reflects structural injustice.}

As these different examples suggest, if the probative value of the evidence rests on structural injustice, this fact serves as an additional reason to recommend relying on that evidence for remedial or supportive uses. When the use is punitive, by contrast, failing the CIP test recommends caution in relying on the evidence. What to say instead in cases in which the probative value of the evidence does not depend on structural injustices?

In the healthcare setting, if the reduced effectiveness of a medical treatment is due to biological factors---say, comorbidities that diminish the treatment’s impact---supportive responses are less appropriate, since they cannot correct the underlying cause of reduced effectiveness. In such cases, the biological cause must be addressed upstream, not through downstream interventions. Similarly, if a camera detects an ongoing crime in a particular area, the immediate objective should be to apprehend the perpetrator and protect the victim. Supportive measures, by themselves, seem inadequate to address the ongoing crime. %By contrast, predictive policing lacks the same urgency. The dangers of over-policing loom larger, so that less punitive, more supportive interventions, such as redesigning public spaces, adding lighting, or organizing community events, seem more apt.  %What explains this divergence in intuitions about the two types of policing algorithms? 

These examples should not be taken to suggest that evidence which passes the CIP test inherently favors punitive responses. Consider the use of blood alcohol concentration (BAC) tests. The biochemical relationship between alcohol intake and BAC levels is grounded in physiology, not structural injustice. Evidence of high alcohol levels can be used punitively---to sanction impaired drivers, suspend licenses, or impose penalties. Yet the same evidence could also support non-punitive interventions, such as initiating detox protocols or offering counseling. CIP does not privilege one use over the other.

In the end, the Principle of Counterfactual Independence (CIP) is best viewed as a filter. When evidence would remain probative in a world without injustice, the principle imposes no additional moral constraints. Absent other moral constraints, such evidence can permissibly be used in either punitive or supportive contexts: to arrest suspects, to stop drunk drivers, to treat alcohol dependency, etc. By contrast, when evidence fails the CIP test---its probative value rests on structural injustice---the principle restricts punitive uses (imposing burdens or sanctions) while at the same time encouraging supportive uses (allocating resources, delivering care, or undertaking remedial interventions).

\bibliographystyle{apalike}
\bibliography{biblio_n3}

\end{document}